\documentclass{PoS}
\newcommand{\be}{\begin{equation}}
\newcommand{\ee}{\end{equation}}
\newcommand{\bea}{\begin{eqnarray}}
\newcommand{\eea}{\end{eqnarray}}
\newcommand{\bee}{\begin{eqnarray*}}
\newcommand{\eee}{\end{eqnarray*}}

\newcommand{\nnu}{\nonumber\\}

\newcommand{\mt}{\tilde m}

\newcommand{\at}{\tilde a}

\def\mt{{\ifmmode\td M_t\else $\td M_t$\fi}}
\def\as{{\ifmmode\alpha_s\else$\alpha_s$\fi}}

\let\td=\tilde

\def\co#1{{\ifmmode{\cal O}_{#1}\else${\cal O}_{#1}$\fi}}
\def\cs#1{{\ifmmode{\cal S}_{#1}\else${\cal S}_{#1}$\fi}}
\def\at{{\ifmmode{\tilde A}\else$\tilde A$\fi}}
\let\nn=\nonumber

\def\fr#1.#2.{{#1\over #2}}

\def\mg{{\ifmmode M_{GUT}\else $M_{GUT}$\fi}}

\newcommand{\s}{\Sigma}
\newcommand{\Sigb}{{\overline\Sigma}}
\newcommand{\oot}{\overline {126}}

\newcommand{\ovl}{\overline}

\title{MSSM Higgs : Window into Susy GUTs}

\ShortTitle{Higgs is the Window}

\author{{Charanjit S. Aulakh}\\
       Department of Physics\\
Indian Institute of Science Education and Research, Mohali,\\
Punjab,140036 India\\
        E-mail: \email{aulakh@iisermohali.ac.in}}

\abstract{The Minimal Supersymmetric SO(10) GUT  has developed
into a fully   realistic theory in which not only are    the gauge
couplings unified    but the known fermion spectrum and mixing
matrices could   fit accurately using the latitude introduced by
inclusion of quantum corrections to the GUT-effective MSSM-SM
matching conditions. The fits  yield predictions about the nature
of the sparticle spectrum on the basis of the required threshold
corrections. This indicated a necessarily large value for $A_0$ in
2008 : well before Higgs discovery at 126 GeV made it a
commonplace assumption.  GUT scale threshold corrections to the
normalization of the  emergent effective  MSSM
  Higgs  ameliorate  the long standing Susy GUT  puzzle of fast
  dimension five operator mediated proton decay. Numerical
   investigation indicates that B-violation rates below or near the current experimental
   upper limits are feasible in fully realistic models.  Our results
    imply that UV completion models with large numbers of fields, like Kaluza-Klein models or String Theory,
    must be able to compute threshold corrections to be
    considered quantitative theories and not  just fables.
     Required improvements in the fitting procedure
  are discussed. A generalization of the NMSGUT by gauging the flavour symmetry of the kinetic
  terms,   while retaining renormalizability and the successful MSGUT  symmetry breaking
  patterns,  may  allow  dynamical generation of the observed Yukawa structure of the MSSM
  via  the spontaneous breaking of the full gauge symmetry down to the MSSM at the unification  scale.
   Focus on the emergence   of the MSSM Higgs from the multiple Higgs doublets in the GUT
  thus   provides a crucial window to view  the energetically remote  UV dynamics specified
    in fully calculable and realistic MSGUTs.}

\FullConference{18th International Conference From the Planck Scale to the Electroweak Scale \\
        25-29 May 2015\\
        Ioannina, Greece }

\begin{document}

\section{Introduction}
 Fundamental Unity of Being has  always loomed large in both Greek(Plotinus) and Indian(Vedanta) Metaphysics.
  Coming  from the  Punjab (fabled in Greece, I was told, since the Alexandrian conquest, as
 $ \pi\epsilon\nu\tau\alpha\pi\omicron\tau\alpha\mu\iota\alpha$)
   which was the meeting ground and melting pot of   Greek and Indian civilizations, it was
 a special  privilege and pleasure to participate in the
 Planck2015 conference at the University of Ioannina and speak on
  Unification : to which Indians and Greeks have alike contributed  much.

 Although it amounts to no more than a simple minded
extension of the Standard model, Grand Unification\cite{patisal}
has stood the test of time as the most convincing scenario for the
unification of forces and types of matter even while  signals
conforming to the expectations it arouses, most notably to proton
decay and Supersymmetry(Susy), remain elusive. Recall that as far
back as 1982 \cite{marcsenj} it was realized that, with a larger
value of the Weinberg angle  than the then current experimental
value ($\sin^2\theta_W \simeq 0.215$) and a top quark(then still
undiscovered) much more massive than the then established limit
($m_{top}> 20 $ GeV), the gauge couplings of Minimal
Supersymmetric SM(MSSM) with a single pair of doublets, but not
the SM nor any other variant of the MSSM, and a Susy breaking
scale in the TeV (up to logarithmic precision), unify accurately
at $M_X^0\simeq 2\times 10^{16}$ GeV. Precision measurements at
LEP eventually confirmed that the Weinberg angle had the required
enhanced value  ($\sin^2\theta_W \simeq 0.23$) and the top quark
was also finally discovered in the range predicted\cite{marcsenj}.
These data also allowed an early appreciation\cite{tbtauni} that
the(tree level) Yukawa couplings of the third generation in the
MSSM  can actually converge at the unification scale $M_X^0$ :
provided the ratio of MSSM Higgs doublet vevs
$<H>/{<\overline{H}>=\tan \,\beta}$ was around $50$. Such a
convergence is to be expected in SO(10) GUTs since they  place an
entire MSSM fermion family \emph{plus} a conjugate neutrino field
($\nu^c_L$) within a single \textbf{16}-plet  of Spin(10).
Spin(10)-less precisely SO(10)- unification further points towards
very small  neutrino masses since the gauge singlet $\nu^c_L$ can
be expected to have a large Majorana  mass from GUT scale vevs,
implying that the effective light(left handed) neutrino masses
will be very small. In the period 1996-1998, when the neutrino
mass magnitudes were still unknown we investigated Minimal
Supersymmetric Left Right models(MSLRMs)\cite{mslrms} with view to
examining their support for  neutrino masses and dark matter. In
these models   ad-hoc R-parity($R_p=(-)^{3(B-L) + 2 S }$)
introduced  for proton longevity in the MSSM is effectively part
of the gauge group since the Abelian factor of the Left Right
gauge group is $U(1)_{B-L}$. When MSLRMs are set up with the SU(2)
triplets required to implement small Seesaw masses via $SU(2)_R
\times U(1)_{B-L}\rightarrow U(1)_{Y}$ breaking it turns out to be
phenomenologically necessary and inevitable that R-parity is
preserved to the lowest energies\cite{mslrms}. Then  the lightest
sparticle (LSP) must be stable, making it an excellent Dark
matter(DM) candidate due to the ``WIMP
miracle''\cite{Wimpmiracle}. This implements\cite{survfittestetc}
a deep connection\cite{pramancsa} between LR symmetry,
Supersymmetry, Neutrino masses, Dark matter and  Susy SO(10) GUTs!
The discovery of the long sought for neutrino oscillations at the
Super-Kamiokande neutrino oscillation/nucleon decay
telescope/microscope\cite{superKdisc}, indicating  milli-eV
neutrino masses, following soon after Top quark discovery, kept up
the drumbeat of discoveries pointing towards Supersymmetric SO(10)
Grand Unification as a   highly natural and favoured framework,
restrictive and ample at the same time, for  Grand Unification of
the known gauge forces and matter fields. Amazingly even the
recent    placement of the SM capstone, namely the discovery of a
SM like Higgs at 125 GeV in 2012,  did not produce any
contradiction with the low energy effective theory being the MSSM
: although by requiring that the soft Susy breaking parameters
($\{m_{\tilde F},A_{0 {\tilde F}}\} \simeq M_{Susy}$) lie in the
Multi-TeV range, it did debunk the `papal' claims that the $M_S$
would only -because of  `naturality' - be of TeV scale.
Interestingly, the detailed  development of a realistic Susy
SO(10) model proposed\cite{aulmoh}  at the beginning of the Susy
Unification story and revived\cite{abmsv} in the SO(10) resurgence
triggered by Super-Kamiokande as the  \emph{Minimal}  (parameter
counting) realistic Susy GUT had already(in 2008\cite{nmsgut})
indicated that $M_{Susy}$ and more specifically the trilinear
parameter $A_0$ would be large (as required by a 125 GeV MSSM
Higgs)! Finally   the determination\cite{theta13L} around 2005
that  minimal SO(10) fits of the fermion mass and mixing data
would only be successful if the PMNS mixing angle $\theta_{13}^L$
-then  quite unknown - was quite large (in the range of 6-12
degrees)  was  confirmed\cite{DayaBay} by the results of Daya Bay
reactor neutrino experiments in 2012.

With this  stunning record of the MSSM and (Minimal Susy)
SO(10)GUT(MSGUT) it was  appropriate to develop the  quantum
structure of the  MSGUT in the full detail possible  due to the
explicit solubility of its spontaneous symmetry
breaking\cite{abmsv} and thereby of the complete superheavy
spectrum\cite{bmsv,fuku,ag2}. Over the last decade these studies
passed significant milestones
\cite{ag2,nmsgut,germblm,blmdm,nstabhedge}. Staying with the
minimal theory without invoking additional discrete symmetries and
Higgs/matter multiplets, and in particular by focussing upon the
implications of retaining only a single pair of light doublets out
of the plethora of doublets generically present in Susy SO(10) and
other realistic(but less economical)  GUTs, it has revealed that
some of the recalcitrant deficiencies  besetting Susy GUTs -like
the lack of constraining/falsifying predictions for the plethora
of soft Susy breaking parameters, the rapidity of dimension 5
mediated proton decay and the lack of insight into the fermion
flavour hierarchy- might also receive new and intriguing answers.
The existence of only a single pair of light Higgs doublets
provides a magic microscope into the innards of the UV completion.
By the nature of their parentage in the multiple Higgs multiplets
of the GUT, the light Higgs are privy to the inner working of the
GUT Higgs dynamics : which is   swept into the observed
renormalizable parameters as far as the other light fields are
concerned. Due to their multiple parentage among the heavy Higgs
multiplets, the superlight(weak scale)  eigenmodes with MSSM Higgs
doublet quantum numbers  can be subject to wave function
renormalization that is strong enough to take the fields close to
``dissolution'' \cite{nstabhedge} i.e. $Z_{H,\overline H}\simeq
0$. Then the canonically renormalized Higgs doublets of the
effective MSSM participate in Yukawa couplings to the matter
fermions which are much enhanced over the values of the tree level
GUT couplings on a ``Higgs dissolution edge'' ( a high dimensional
sub-manifold of the multidimensional GUT superpotential parameter
space where $Z_{H,\overline H}\simeq 0$).   The   GUT matter
Yukawa couplings required to account for the observed fermion
masses are thus much diminished. Since they also    enter
quadratically in the proton decay amplitudes it follows that the
proton lifetime is   much enhanced for realistic SO(10) Yukawa
couplings. Thus `doublet-triplet' splitting, long regarded as the
bane of GUTs, actually seems to resolve their longstanding
\cite{weinbgdim5} structural difficulties with dimension five
operators by  a completely novel and generic mechanism. Finally,
given that a completely acceptable UV completion of the MSSM may
be  available in the NMSGUT, we asked whether the sublation of the
MSSM data in the NMSGUT is not the optimal starting point to
attack the problem of flavour. Dynamical generation of the
successful MSGUT matter couplings by promotion of the flavour
symmetry to a gauged $O(N_g)$  symmetry broken by the MSSM GUTs
(promoted to irreps of the flavour symmetry) and \emph{still}
under the crucial constraint that a single pair of Higgs doublets
remain light in the effective MSSM is
feasible\cite{yumguts,bajcmelfo}. Surprisingly, the fermion
hierarchies generated  tend to carry the features we associate
with the MSSM : small CKM and large PMNS mixings, hierarchical
fermion Yukawas and so on. In this proceeding  we run through the
main issues of our current work these two areas. Details may be
found in \cite{nmsgut,nstabhedge,yumguts,bajcmelfo}.

\section {Minimal Susy SO(10) Grand Unified Model}

 The  model in question was introduced long ago\cite{aulmoh}. Besides the 3
   Spin(10)  \textbf{{16}}-plets containing matter fields
 and the  \textbf{{45}}-plet of gauge vector multiplets the  MSGUT/NMSGUT
 utilizes antisymmetric
 $  {{\Phi_{[ijkl]}(210),\Sigma_{[ijklm ]}(126),\overline
 \Sigma_{[ijklm}](\oot)}}$  as Higgs multiplets
 to break the Susy SO(10) symmetry in one step (as is necessary  to avoid  pseudo-goldstone
 problems\cite{aulmoh,rparso10}) to the MSSM. In addition to these
 ``AM'' multiplets  which perform the GUT symmetry breaking and
   generate neutrino masses($ \mathbf{{\oot}}$),
 one has also the ``FM'' multiplets $\textbf{10,120}$ responsible
 only for charged fermion masses. Interestingly this set makes up the
 full complement of antisymmetric irreps  possible in SO(10) !
 The  superpotential  is built from mass terms (SO(10)
 contractions are indicated by  c-dots or powers)
 \bea
 m: {\bf{210}}^{\bf{2}} \quad ;\quad M : {\bf{126\cdot{\overline {126}}}}
\quad ;\quad M_H : {\bf{10}}^{\bf{2}}\quad;\quad m_{\Theta}
:{\bf{120}}^{\bf{2}} \eea

and trilinear couplings  :
 \bea
 \lambda &:& {\bf{210}}^{\bf{3}} \qquad ; \qquad \eta :
 {\bf{210\cdot 126\cdot{\overline {126}}}}
 \qquad;\qquad \rho :{\bf{120\cdot 120 \cdot{ { 210}}}}
\qquad;\qquad k : {\bf{ 10\cdot 120\cdot{ {210}}}} \nnu &&\gamma
\oplus {\bar\gamma} : {\bf{10 \cdot 210}\cdot(126 \oplus{\overline
{126}}})\qquad;\qquad \zeta \oplus {\bar\zeta}  :  {\bf{120 \cdot
210}\cdot(126 \oplus {\overline {126}}})\nnu && {\bf{16}}_A\cdot
{\bf{16}}_B \cdot(h_{AB} {\bf{10}} + f_{AB}{\bf{\overline {126} }}
+g_{AB} \bf{120} )
 \eea
 The couplings $h,f(g)$ are complex (anti)-symmetric matrices   flavour
 space  because of  the properties of the SO(10) Clifford algebra.
  Either  $h$ or $f$  can be  chosen   real and diagonal  using the $U(3)$ flavor symmetry
   of the matter kinetic terms. Five   phases  say  of
 $m,M, \lambda ,\gamma,\bar\gamma$ may be  set to zero by phase
 conventions.
 Matter  Yukawa couplings  contain 21 real
parameters. To implement the  crucial ``one light Higgs pair ''
consistency condition (a.k.a. doublet-triplet splitting by fine
tuning) $M_H$  is fine tuned to ensure that the  $6\times 6 $
Higgs doublet mass matrix $\mathbf{\cal{H}}$ has zero
determinant(and thus a pair of  null left and right eigenvectors)
 keeping two Higgs doublets  ($H,{\overline{H}}$) of the
effective MSSM light. This   leaves  23 magnitudes and 15 phases
as parameters.    $H,{\overline{H}}$  are a mixture  of the (6
pairs of the MSSM type)  doublet fields in the GUT. The mixture
parameters(`` Higgs fractions''  $\alpha_a,\bar\alpha_a,a=1..6$
\cite{abmsv,ag2}) are functions of the NMSGUT superpotential
parameters which enter  the left and right null eigenvectors of
$\mathbf{\cal{H}}$. The symmetry breaking vevs  \bea
 {\langle(15,1,1)\rangle}_{210}& : & a
  \qquad \langle(15,1,3)\rangle_{210} : w \quad
\quad\langle(1,1,1)\rangle_{210}~ :  ~ p \nn\\
  \langle(10,1,3)\rangle_{\oot} ~& :&
    \bar\sigma \quad  ;\quad
\langle({\overline{10}},1,3)\rangle_{126} ~ :
 \sigma   \nn\eea
preserve SUSY from  D Terms violation by
$|\sigma|=|{\overline{\sigma}}| $, while the 4 coupled F term
equations $F_{a,p,w,\sigma/\bar\sigma}=0$       are then
analytically soluble\cite{abmsv}! The GUT scale vevs( in units of
$m/\lambda$)  are known  functions of a complex varible  $x$ which
solves  the cubic ($\xi ={{ \lambda M}\over {\eta m}} $) \be 8 x^3
- 15 x^2 + 14 x -3 +\xi (1-x)^2=0 \label{cubic} \ee

592 Higgs Chiral  and 33 Majorana heavy gauge supermultiplets
occur in 22 complex (pairs) and 4 real MSSM representation types.
Explicit solution  of SSB allows explicit determination of their
mass matrices and eigenvalues. The complete spectrum of the 26
different MSSM irrep types and tree level effective Superpotential
is available \cite{nmsgut} and extends the    MSGUT
result\cite{ag1,ag2,bmsv,fuku}. This allows calculation of the
superheavy threshold correction to the gauge and MSSM Yukawa
couplings due to circulation of super heavy fields in the self
energy diagrams for the light fields and the consequent finite
renormalizations.

From the point of view of the consistency of the GU   picture, the
mass matrix ${\cal H}$ of the six pairs of MSSM type doublet
irreps carrying $[1,2,\pm 1]$ of $SU(3)\times SU(2)_L\times
U(1)_Y$    is   most crucial. These doublets come from \emph{all}
the Higgs representations present in the theory and mix together
thoroughly. As a result of being woven so intimately into the
innards of the superheavy spectra and couplings and yet being
constrained by the necessity that   Supersymmetric gauge
unification   yield the  super light Higgs doublet  pair   which
is the \emph{sine qua non} of Susy unification, the light MSSM
Higgs provides a unique vantage point to view the inner workings
of the GUT. \emph{In other words the composition of the Higgs is
the portal of irruption of the superheavy GUT world into the our
low energy reality.} Pursuing this line of thought further we are
led to a detailed investigation of the effects of superheavy
fields on the couplings of the superlight fields.
\section{GUT scale threshold corrections and Baryon decay rate}
 Superpotential parameters renormalize only by
wave function corrections. This implies\cite{weinberghallwright}
that the threshold corrections to the matching conditions between
gauge and Yukawa couplings can be calculated simply by considering
the (finite) wavefunction renormalization of the light fields due
to loops containing (at least) one heavy supermultiplet. Since the
complete decomposition of the NMSGUT trilinear vertices by MSSM
quantum numbers was already obtained\cite{ag2,nmsgut} when
calculating the mass spectra,  the one loop effects are
straightforward to compute. Threshold corrections to the gauge
coupling matching conditions depend upon the ratios of masses.
There is a freedom to choose $M_X$ within a small range around the
MSSM unification scale $M_X^0=2\times 10^{16}$ GeV as well as the
SO(10) coupling $\alpha_G(M_X)$. The spread of mass eigenvalues
allows cancellation among threshold corrections and thus a
sensible result: belying\cite{ag2} the expectation\cite{dixitsher}
that such corrections make Susy SO(10) unification meaningless.

The threshold corrections to the Yukawa couplings of the MSSM are
much more tedious and run\cite{nstabhedge} to some 1500 sums  each
of which runs over one or more  multiplicity indices of the 26
 superheavy MSSM irrep types which mix.

For a generic light field  $\Phi_i$ the  one loop wave function
constants  ($Z= 1 -{\cal K}$) have form  :\bea{\cal K}_i^j=-
{\frac{g_{10}^2}{8 \pi^2}} \sum_{\alpha,k}{Q^\alpha_{ik}}^*
{Q^\alpha_{kj}} F(m_\alpha,m_k) +{\frac{1}{32 \pi^2}}\sum_{kl}
Y_{ikl} Y_{jkl}^* F(m_k,m_l)\label{threshcor} \eea
 here $A_\mu^\alpha$  is a   SO(10) \emph{heavy} gauge boson
$Q^\alpha$ the associated generator ( $g_{10}$ is the  SO(10)
 gauge  coupling. $m_{\alpha,k}$ are
heavy gauge and chiral multiplet masses). The generic Yukawa
couplings are defined by the superpotential $W={\frac{1}{6}}
Y_{ijk}\Phi_i \Phi_j \Phi_k$. The sums in eqn. (\ref{threshcor})
run over pairs of intra-loop fields with at least one member
superheavy and $F(m_\alpha,m_k)$ is a standard 1-loop
Passarino-Veltman function: \bea F_{12}(M_A,M_B,Q)={1\over
{(M_A^2- M_B^2)}}( M_A^2\ln {M_A^2\over Q^2} -M_B^2\ln {M_B^2\over
Q^2} )- 1 \eea which reduces to just $
F_{11}(M_A,Q)=F_{12}(M_A,0,Q)=  \ln {M_A^2\over Q^2} - 1 $ when
one field is light ($M_B\rightarrow 0)$.

The simplest   contribution to a Higgs line correction is in the $
W[6, 3, 2/3 ]\otimes \bar Y [(6, 2, 13 ]$ channel(see \cite{ag2}
for the alphabetical notation for MSSM irreps which occur in the
MSGUT) :\bea
  K_{W\bar Y} &=&\biggr|- \gamma U^H_ {11} + \frac{2\eta}{ \sqrt{3}} U^H_
  {21}- \zeta U^H_ {51} +\frac{i\zeta}{\sqrt{3}} U^
  H_ {61}\biggr|^2 F_{12} (m^W,m^Y,Q)
  \eea
Here $U^H$ are matrices that participate in the diagonalization of
the Higgs doublet mass matrices. When multiple copies of the
running heavy multiplets mix   there are diagonalization matrices
for those mass matrices present as well. An example of a gauge
correction (for the matter light line $\bar u$, $m_\lambda$ are
gaugino masses) is \bea
  ({ {16\pi^2}}) {{\cal K}}^{\bar{u}}&=& - 2 {g_{10}^2}
(0.05 F_{11}(m_{\lambda_G},Q) + F_{11}(m_{\lambda_J},Q) +
F_{11}(m_{\lambda_F},Q) +4 F_{11}(m_{\lambda_X},Q) \nnu &&+2
F_{11}(m_{\lambda_E},Q))\eea

The (over 1500) other terms in the the dressing of the light lines
entering an MSSM vertex are considerably more involved and we
refer the reader to the original work  for   details
details\cite{nstabhedge}.

 The dressing of light by heavy lines  implies \cite{weinberghallwright} a finite wave
function renormalization in the fermion and Higgs Kinetic terms
\bea {\cal L}=[\sum_{A,B}( {\bar f}_A^\dagger (Z_{\bar f})_A^B
{\bar f}_B +{f}_A^\dagger (Z_{f})_A^B {f}_B ) + H^\dagger Z_H H +
{\ovl H}^\dagger Z_{\ovl H} {\ovl H}]_D +..\eea

Diagonalizing  the matter wavefunction dressing matrix by Unitary
matrices   $U_{Z_f},{\ovl U}_{Z_{\bar f}}$ ($U^\dagger Z U=
\Lambda_Z$), one defines a  new basis  to put the light Kinetic
terms in canonical form   So the  MSSM Yukawas match the dressed
Yukawa couplings   of the   new (canonical Kinetic term) light
fields $\tilde Y_f$: \be\tilde Y_f= \Lambda_{Z_{\bar
f}}^{-\frac{1}{2}} U_{Z_{\bar f}}^T {\frac{Y_f}{\sqrt{Z_{H_f}}}}
U_{Z_f} \Lambda_{Z_f}^{-\frac{1}{2}} = \tilde{U}_{Z_{\bar f}}^T
{\frac{Y_f}{\sqrt{Z_{H_f}}}} \tilde{U}_{Z_f} \label{Ynutilde}\ee
 and not to the original SO(10) tree level ones.

 There are precisely 26 different combinations of
the 26 MSSM representation  types that  run in the loops on the
Higgs lines in the MSSM matter fermion Yukawa vertices. The SO(10)
Yukawa couplings $(h,f,g)_{AB}$ which enter the light field Yukawa
couplings to the light Higgs  also enter   the coefficients
 of the $d=5 $ baryon decay operators in the
effective dimension 4 superpotential obtained by integrating out
the heavy chiral supermultiplets that mediate baryon
decay\cite{ag1,ag2,nmsgut}. These operators do not have external
light Higgs lines and as a result are \emph{not} boosted by the
Higgs dressing. Thus suppression of matter field Yukawas by Higgs
on the dissolution edge($Z_{H,\bar H}\simeq 0$) self-consistently
lowers the SO(10) matter Yukawas ($ \{f,g,h\}_{AB}$)  causing
their dimension 5 proton decay effect to be much suppressed while
simultaneously ensuring that the dressing of matter fields by
heavy fields is ultraweak and therefore no enhancement of Baryon
violation due to matter field dressing.  It is an interesting and
open question as to whether higher dimensional superexotic (B-L
violating etc) operators \emph{containing } a light Higgs field
can receive a boost to observable levels due to Higgs wave
function enhancement. Such operators are specially interesting
from the point of view of novel scenarios of Baryogenesis and B
violation\cite{babmoh} except that they tend to be badly
suppressed by the additional powers of $M_X^{-1}$ incurred due to
their higher dimension. The light field  dressing also has
effects on the soft breaking terms  that may eventually prove
important in precision fits of MSSM data in terms of GUT
couplings.

Using these corrections we were able to show \cite{nstabhedge}
that   \emph{prima facie}  a solution of the dimension 5 proton
decay problem\cite{weinbgdim5}, outstanding for long, was feasible
in a realistic Susy GUT. This was done using an iterated fitting
at high scales(of the RG extrapolated low energy Yukawa couplings
of the MSSM) and at low scales of the SM Yukawas in terms of the
Susy threshold corrected GUT generated Yukawas. Large $\tan \beta$
driven (H-Hbar mixing) threshold corrections to down type fermion
yukawas prove crucial towards achieving realistic down type quark
masses in the first and second generations.  We  can fit charged
fermion masses   $y_t\simeq y_b\simeq y_\tau(M_X)$ and
$\tan\beta\simeq 50$  if   MSSM radiative corrections raise $
Y^{GUT}_{d,s}$ by 3-4 times while $ Y^{GUT}_{b}$ is  lowered by
around  $5\%$. GUT scale threshold corrections which repair fast
proton decay rate \emph{also} loosen stringent tree level
constraints\cite{ten120cnstrts} on $10\oplus 120$ generated
charged fermion masses. The iterative procedure has  drawbacks
such as occurrence of large SO(10) gauge couplings and ambiguities
regarding
 which  couplings to use at successive stages. Moreover the
light sparticle masses remained  to be loop corrected on the fly.
These improvements have been incorporated and a non-iterative
search for fits, based upon a  random choice in the large ($>44$)
dimensional parameter space followed by threshold corrections and
running all quantities down to the Electroweak scale,  before
trying the  fit  the known low energy data,  is underway, but
exhibits slow convergence due to  the plethora of local minima
encountered by the Nelder-Mead amoeba as it traverses the high
dimensional parameter landscape. Calculation of the  off diagonal
1-loop corrections in the MSSM, generalizing \cite{baggerpierce}
are also required for a fully satisfactory treatment and are
underway. Our searches show it is difficult to suppress the
coefficient of the   dimension 4 B-violation operator in the
effective superpotential below about $10^{-21.5}$ GeV${}^{-1}$.
The resultant value of the proton lifetime also depends on the
soft susy spectrum, so that it may be necessary\cite{bps} to
optimize both GUT scale and $O(M_{susy})$ parameters to achieve
acceptable B-violation. This case has the upside that   fully
realistic fits will both constrain Soft susy couplings and predict
a proton life time upper bound near to the current limits. Minimal
Susy SO(10) may thus prefigure an upcoming discovery of proton
decay.

 A comprehensive, neat and predictive fit or else a falsification
  of the NMSGUT as it stands is thus feasible and on the horizon.
   At this stage, rather than  any
particular fit  we wish to emphasize  that this serious treatment
of the quantum dynamics of the only fully realistic and calculable
GUT available has made it clear  that \emph{any} serious UV
completion must be able to carry out a corresponding estimation of
heavy field effects on light field couplings or abandon any
pretense at a quantitatively valid unification. Since several
pre-eminent types of Unification  prone to making  ambitious
claims are yet far from even being able to consistently specify a
separation of light from heavy modes, leave alone their dressing
in terms of heavy mode quantum effects \emph{separately} from
light field ones, it is clear that the bar for candidate models
has significantly been raised by our  studies.

\section{Yukawon Flavour Models }

 The  (MS)SM Fermion kinetic terms carry a large global symmetry($U(3)_Q\otimes U(3)_u\otimes
U(3)_d\otimes U(3)_L\otimes U(3)_l $) broken only by the
 Yukawa couplings. To understand the origin of the observed
  flavour hierarchy, it is natural to ask whether
 the breaking terms are not the  residue of  spontaneous
 breaking of   flavour symmetry by some unknown high scale
 dynamics. Such ``Yukawon'' models based on the SM/MSSM  must normally live with the
 uncomfortable necessity of  \emph{non-renormalizable} dynamics
 since the Yukawa terms are already of dimension 4 and the
 couplings promoted to scalar field vevs raise the operator from
 marginal to irrelevant. Model building
 with recourse to   generation of structurally crucial  marginal
 operators via  irrelevant operators is always arbitrary and
  unconvincing except when directly  motivated by
 phenomenology. One should rather first
 ask whether there is any sign of of flavour unification in
 RG evolution into the UV. The only known hint available is the
 GUT scale \cite{tbtauni} unification of third generation couplings in the high $\tan\beta$
 MSSM motivated by  Susy SO(10) GUTs! Thus it is natural to ask
 whether the GUT Yukawa couplings found  in   NMSGUT fits of all the  SM data
 might arise from field vevs. In such  NMSGUT  based models, however, the Higgs and
 Yukawon functions can be comfortably combined,
  \emph{without sacrificing renormalizablity},   and the flavour
 symmetry broken at the only scale where  there \emph{is} a hint of flavour
 unification : the GUT scale.  In \cite{yumguts,bajcmelfo} we
showed that this economical scenario is actually realizable in the
context of  models where a \emph{gauged} $O(N_g)$ flavour symmetry
is appended to the NMSGUT without disturbing the structure of
NMSGUT apart from the  promotion of Higgs from bland to flavoured
and the inverse demotion of $SO(10)$ matter  Yukawas from
flavoured to to bland. We note that the freedom to  gauge flavour
achieved in these models is  in line with the philosophy we have
consistently abided by, first in MSLRMs\cite{mslrms} and then in
MSGUTs\cite{abmsv,ag2,nmsgut,blmdm,nstabhedge} : invoke no ad hoc
global symmetries whether continuous or discrete and shelter
always under the sole reliable dynamical  principle revealed by
20th century physics : local gauge invariance.

Although the global symmetry of the Spin(10) GUT matter  kinetic
terms is $U(N_g)$, gauging just an   $O(N_g)$
 subgroup seems  workable. Gauging this symmetry can ensure that no Goldstone bosons
  arise when it is spontaneously broken.  With    a unitary family
 gauge symmetry,   complex representations  introduce
   anomalies and require  irrep doubling   to cancel gauge  anomalies and to allow
   formation of  holomorphic  invariants
    for the  superpotential.  Moreover one finds that
 a Unitary  flavour gauge group  implies   half the
  effective MSSM  matter Yukawa couplings vanish. An $O(N_g)$
  flavour gauge group is free   of these defects
We emphasize  that in contrast with previous models
 (see e.g. \cite{koide}) our  model  is renormalizable and    GUT based.

 The Yukawon GUT superpotential   has   the MSGUT  form    :
  \bea W_{GUT}&=&\mathrm{Tr}( m \Phi^2 + \lambda \Phi^3 + M \Sigb .\s +\eta \Phi .\Sigb.
  \s)
  +  \Phi.H.(\gamma \Sigma +\bar\gamma  . \Sigb) + { M_H}
  H.H)\label{WGUT} \nnu
  W_F &=&\Psi_A .((h H) + (f \s) + (g\Theta) )_{AB}\Psi_B \label{WF}\eea

The manner of insertion of the  ${\bf{120}}$-plet is  indicated in
$W_F$ but so far we have  have studied only MSGUTs (i.e with
$\mathbf{10,\oot}$). Anyway\textbf{120}-plet does not  take part
in GUT scale symmetry breaking since it has no SM  singlets.  The
main novelty  is that  MSGUT Higgs fields now carry a symmetric
  representation of the $O(N_g)$  family symmetry :
  $\{\Phi,\Sigb,\s,H \}_{AB}; A,B=1,2..N_g$ (  the
  matter $\mathbf{16}$-plets $\psi_A$ are vector $N_g$-plets). Now the
  couplings $h,f,g$ are no longer matrices but
just (complex) numbers while the Yukawons
  carry symmetric ($\{H,\Sigb\}_{AB}= \{H,\Sigb\}_{BA}$)
   and anti-symmetric ( $\Theta_{AB}= -\Theta_{BA} $ )
  representations  of   $O(N_g)$ as required by the properties
   of Spin(10) 16-plet bilinears. Thus, for $N_g=3$, the number of
   (real) matter  Yukawa parameters come  down from 15
    ($Re[h_{AA}],f_{AB})$   to just
    3 ($Re[h],f$) without the \textbf{120}-plet
 (6 additional to just 2 additional  with the ${\mathbf{120}}$-plet).
 Such  renormalizable flavour unified ``Yukawon'' GUTs  should thus  be called Yukawon Ultra-Minimal
    GUTs(\textbf{YUMGUT}s).

The spontaneous symmetry breaking at the GUT scale due to the
above superpotential is largely analogous to the MSGUT case
although  algebraic complications require numerical solution of
the F term conditions\cite{yumguts}. The contribution of the MSGUT
type Higgs fields to the flavour group D terms cannot vanish and
this  breaks supersymmetry unless additional chiral multiplets
charged under flavour but not under SO(10) and  free to cancel the
MSGUT type Higgs contribution in the flavour D-Terms are
introduced. Remarkably this \emph{can} be done\cite{bajcmelfo}
using ``Bajc-Melfo'' type metastable supersymmetry breaking which
-when generalized  by flavour gauging- leaves some Chiral
multiplet vevs free to be determined by the flavour D term
conditions. Thus a remarkable connection between spontaneously
broken  gauged flavour and supersymmetry breaking emerges. This
strange and novel relationship  in the UV theory leaves a clear
relic in the effective theory consisting of very light `moduli'
type fields(the fermionic components get masses only radiatively
while Bosonic components get the  usual soft susy breaking scale
masses) which may be very light(sub-GeV to sub-Electro-weak)  Dark
matter candidates of a novel type not ordinarily found in SO(10)
GUTs.
 In these models  the hard parameters of the effective MSSM
 and the soft supersymmetry
breaking parameters are  determined by the two parameters of the
hidden sector superpotential and  the Planck scale together with
the (reduced) parameter space of the NMSGUT sector
superpotential.. What we wish to emphasize here is that the
constraint $Det[{\cal{H}}]=0$ imposed on the generalization of the
$4\times 4$ doublet matrix in the MSGUT to a $2N_g(N_g+1)$
dimensional Higgs mass matrix in the YUMGUT and, correspondingly,
of the $6\times 6$ doublet matrix in the NMSGUT to a $N_g\times
(3N_g+1)$ dimensional matrix is sufficient to generate solutions
with all the observed features of the fermion hierarchy.
See\cite{yumguts,bajcmelfo} for details.

\section{Discussion}
In this talk we   focussed on the crucial role of the consistency
condition which ensures  a single pair of Light Higgs doublets in
the effective MSSM arising from the fully calculable Minimal
Supersymmetric GUT introduced long ago\cite{aulmoh} and developed
over the years into a fully consistent theory. We have shown how
the calculability of quantum effects not only allows
implementation   of the fermion hierarchy and fitting of the full
MSSM gauge and fermion data data, including neutrinos, but also
ameliorates  the longstanding and generic problem\cite{weinbgdim5}
with dimension 5 B violation operators  by showing an easy
approach via inescapable quantum effects (even with small
couplings, due to the large number of heavy fields) to the ``Higgs
dissolution edge'' in parameter space. On this  "edge"  the
wavefunction renormalization constants of the light Higgs
multiplets arising from the NMSGUT nearly vanish, thus amplifying
operators (like the MSSM Yukawa couplings) where Higgs fields
enter while suppressing the troublesome B-violation operators by
several orders of magnitude. The full implications of this
mechanism, both for the achievable $d=5$ B-violation operator
rates and for higher dimension rates are still under
investigation. However it is likely that fully consistent  fits,
if found, will still predict proton lifetimes close to current
lower limits. Thus  the NMSGUT indicates proton decay may be
visible in the next generation of neutrino tele-/nucleon micro-
scope detectors currently under construction or planning.

\bibliography{references}

\end{document}